\UseRawInputEncoding
\documentclass[11pt]{article}
\usepackage{amsmath}
\usepackage{amsthm}
\usepackage{amsfonts}
\usepackage{bbm}
\usepackage{mathrsfs}
\usepackage{graphicx}
\usepackage{sectsty}
\RequirePackage{setspace}
\usepackage{appendix}
\usepackage{float}
\usepackage{url}
\usepackage{caption}
\usepackage{lineno}
\usepackage[round]{natbib}
\usepackage[table,xcdraw]{xcolor}
\usepackage{float}

\newcommand{\captionfonts}{\footnotesize}
\makeatletter 
\long\def\@makecaption#1#2{%
\vskip\abovecaptionskip
\sbox\@tempboxa{{\captionfonts #1: #2}}%
\ifdim \wd\@tempboxa >\hsize
{\captionfonts #1: #2\par}
\else
\hbox to\hsize{\hfil\box\@tempboxa\hfil}%
\fi
\vskip\belowcaptionskip}
\makeatother 

\begin{document}

\title{\bf Bell's Inequalities and Entanglement \\ in Corpora of Italian Language}

\author{Diederik Aerts$^*$, Suzette Geriente\footnote{Center Leo Apostel for Interdisciplinary Studies, 
        Vrije Universiteit Brussel (VUB), Pleinlaan 2,
         1050 Brussels, Belgium; email addresses: diraerts@vub.be,sgeriente83@yahoo.com} 
        , Roberto Leporini\footnote{
        Department of Economics, University of Bergamo, via dei Caniana 2, Bergamo, 24127, Italy; email address: roberto.leporini@unibg.it}
        $\,$ and Sandro Sozzo\footnote{Department of Humanities and Cultural Heritage (DIUM) and Centre CQSCS, University of Udine, Vicolo Florio 2/b, 33100 Udine, Italy; email address: sandro.sozzo@uniud.it}              }
\date{}

\maketitle

\begin{abstract}
\noindent 
We analyze the results of three information retrieval tests on conceptual combinations that we have recently performed using corpora of Italian language. The tests have the form of a `Bell-type test' and were aimed at identifying `quantum entanglement' in the combination, or composition, of two concepts. In the first two tests, we studied the Italian translation of the combination {\it The Animal Acts}, while in the third test, we studied the Italian translation of the combination {\it The Animal eats the Food}. We found a significant violation of Bell's inequalities in all tests. Empirical patterns confirm the results obtained with corpora of English language, which indicates the existence of deep structures in concept formation that are language-independent. The systematic violation of Bell's inequalities suggests the presence of entanglement and, indeed, we elaborate here a `quantum model in Hilbert space' for the collected data. This investigation supports our theoretical hypothesis about entanglement as a phenomenon of `contextual updating', independently of the nature, micro-physical or conceptual linguistic, of the entities involved. Finally, these findings allow us to further clarify the mutual relationships between entanglement, Cirel'son's bound, and no-signaling in Bell-type situations.
\end{abstract}
\medskip
{{\bf Keywords}: concept combinations, linguistic corpora, Bell's inequalities, quantum modeling, entanglement}

\section{Introduction\label{intro}}
The year 1935 saw the publication of two fundamental articles which illustrated the peculiarities of `quantum entanglement'. In one of these articles, Albert Einstein, Boris Podolsky and Nathan Rosen (EPR) showed that, whenever a composite quantum system, or `entity', made up of two individual entities, is in an entangled state, the component entities exhibit a specific type of statistical correlations, known as `EPR correlations' \citep{epr1935}. In the other article, Erwin Schr\"{o}dinger showed that two entangled quantum entities, though separated in space, behave as if they were actually `non-separated' \citep{schrodinger1935}. Then, in 1951, David Bohm introduced the archetypical situation of a composite bipartite quantum entity made up of two spin 1/2 quantum entities which fly apart when the composite entity is in the singlet spin state \citep{bohm1951}. If one imagines that, (i) if one spin is forced ``up'' by the measurement apparatus applied to it, then, as a consequence, the other spin is `immediately' forced ``down'', even when no measurement is performed on it, and (ii) this process is independent of the distance between the two quantum entities, the phenomenon reveals a sort of `spooky action at-a-distance', a locution coined by Einstein to stress the strange aspects of entanglement. Next, in 1964, John Bell put forward the idea of how an empirical test can be designed to detect the presence of entanglement in physical domains. More precisely, what Bell did was deriving an inequality that should not be violated under the assumption, reasonable in classical physics, of `local separability', or `local realism', whereas the inequality is violated in quantum mechanics \citep{bell1964}. After the seminal article by Bell, it became clear that, due to entanglement, quantum entities exhibit genuinely non-classical aspects, such as `contextuality', `non-separability', and `non-locality'. Bell's work was also relevant for the so-called `hidden variables program', initiated after the EPR article, because these results entail that any hidden variables completion of quantum mechanics must be non-local. Finally, one can prove that the EPR correlations cannot be modeled in a classical Kolmogorovian probability framework \citep{accardifedullo1982,pitowsky1989}.

In the last two decades, research on entanglement has flourished in physics. Indeed, at a theoretical level, several variants of Bell's original inequality have been derived, now generally called `Bell's inequalities' (see, e.g., \citet{horodecki2009,brunner2014}). At an empirical level, several tests on micro-physical entities have been performed which confirm the predictions of quantum mechanics (see, e.g., \citet{genovese2005,vienna2013,urbana2013}). Among Bell's inequalities, the `Clauser--Horne--Shimony--Holt (CHSH) inequality' \citep{chsh1969} is particularly suited, not only for empirical tests of non-locality in quantum physics, but also as a test to detect the presence of entanglement, within and beyond physical domains. 

Concerning the latter point, several theoretical and empirical studies have appeared that investigate the presence of entanglement in conceptual-linguistic domains, mainly in the combination of two concepts (see, e.g., \citet{bruza2009,bruzakittonelsonmcevoy2009,aertssozzo2011,aertssozzo2014,gronchistrambini2017,arguelles2018,aertsetal2019,
beltrangeriente2019,arguellessozzo2020,aertsbeltrangerientesozzo2021,aertsarguellesbeltrangerientesozzo2023b,bertinietal2023,
aertsleporinisozzo2025} and references therein). This investigation fits a growing research program, the `quantum cognition program', that applies the mathematical formalism of quantum mechanics in the modeling of high-level cognitive processes, such as perception, categorization, language, judgement, and decision-making (see, e.g., \citet{aerts2009a,khrennikov2010,piwowarskietal2010,busemeyerbruza2012,aertsbroekaertgaborasozzo2013,aertsgaborasozzo2013,havenkhrennikov2013,dallachiaragiuntininegri2015a,
dallachiaragiuntininegri2015b,kwampleskacbusemeyer2015,melucci2015,aertsetal2018,haven2018,pisanosozzo2020,aertssassolisozzoveloz2021} and references therein).
 In particular, some of us have put forward a `realistic-operational approach' for human concepts and their combinations.\footnote{Throughout this article, we write concepts using capital letters and italics, e.g., {\it Animal}, {\it Fruit}, {\it Vegetable}, etc. This is often the way to refer to concepts in cognitive psychology.}  In it, a concept is an entity in a defined state that incorporates the meaning of the concept and can change under the influence of a context \citep{aertssassolisozzo2016}. The realistic-operational approach enables the modeling of conceptual entities using the mathematical formalism of quantum mechanics in Hilbert space \citep{aerts2009a,aertsgaborasozzo2013,pisanosozzo2020,aertssassolisozzoveloz2021}.

Regarding specifically the identification of entanglement in conceptual combinations, we have performed various `Bell-type tests' in which we have tested the presence of entanglement by means of a violation of the CHSH inequality. These tests include text-based \citep{aertssozzo2011,aertssozzo2014,aertsarguellesbeltrangerientesozzo2023b,bertinietal2023} and video-based \citep{aertsleporinisozzo2025} cognitive tests on human participants, information retrieval tests on corpora of documents of English language \citep{beltrangeriente2019,aertsbeltrangerientesozzo2021}, and image retrieval tests on web search engines \citep{arguelles2018,arguellessozzo2020}. We have particularly tested the combination {\it The Animal Acts}, meant as a composite bipartite conceptual entity made up of the individual conceptual entities {\it Animal} and {\it Acts}, where the term ``acts'' refers to the action of emitting a sound by the animal. We have also tested the combination {\it The Animal eats the Food}, meant as a composite bipartite conceptual entity made up of individual conceptual entities {\it Animal} and {\it Food} (see Section \ref{empirical}).
 
While all Bell-type tests significantly violated the CHSH inequality, thus indicating the presence of entanglement in conceptual-linguistic domains too, similarly to how the violation of the CHSH inequality in physical domains is provoked by entanglement, we also found a systematic violation of the marginal law conditions and a frequent violation of the CHSH inequality beyond the so-called `Cirel'son's bound' \citep{cirelson1980,cirelson1993} (see again Section \ref{empirical}). These two results were unexpected from the point of view of quantum physics. This led us to work out a new theoretical perspective to rigorously characterize entanglement in all its aspects of non-classicality \citep{aertsetal2019,aertsleporinisozzo2025}. While the new theoretical perspective has many analogies with some established results on physical entanglement, we also identified some points where the two perspectives, physical and conceptual-linguistic, diverge, in particular with respect to the relationship between the violation of the marginal law conditions and the possibility of `signaling' (see Section \ref{theoretical}).

In this article, we deepen and extend the investigation above and present the results of three new information retrieval tests we have recently performed on the conceptual combinations {\it The Animal Acts} and {\it The Animal eats the Food}, using selected corpora of Italian language. The results strongly confirm the empirical patterns identified in Bell-type information retrieval tests on corpora of English language. This indicates that we have identified some general and deep structures underlying conceptual entities and their formation/combination which are independent of the specific language, English or Italian, that is used to reveal them. In particular, we show that the CHSH inequality is significantly violated in all tests, which reveals the presence of entanglement between the component concepts (see Section \ref{tests}).

Next, we apply the quantum-theoretic framework that enables modeling of any Bell-type test and represent collected data in Hilbert space, showing that entanglement occurs at both state and measurement levels, hence it is stronger than the entanglement that is typically detected in quantum physics tests. More important, the modeling reveals that entanglement is the way to formally express `meaning'. Equivalently, both concepts {\it Animal}  and {\it Acts} carry meaning, but also the combination {\it The Animal Acts} carries its own meaning, and this meaning is not simply related to the separate meanings of {\it Animal} and {\it Acts} as prescribed by a classical compositional semantics. It also contains emergent meaning almost completely caused by its interaction with the wide overall context (see Section \ref{model}). 

In recent work, we have called `contextual updating’ the complex process by which meaning is attributed to the combined concept, and it occurs at the level of entanglement formation \citep{aertsarguellesbeltrangerientesozzo2023a}. We believe that this is the general way to express entanglement in physical 
domains too, and is more convincing than the typical `spooky action at-a-distance' view through which physicists tend to understand the phenomenon of entanglement (see Section \ref{conclusion}).

\section{Detection of entanglement in physical and conceptual-linguistic domains\label{empirical}}
We review here the empirical setting that is typically used to detect entanglement in both physical and conceptual-linguistic domains. This detection involves testing one of Bell's inequalities, namely, the `CHSH inequality' \citep{chsh1969}.

An empirical test for the detection of entanglement in micro-physical domains is called a `Bell-type test' and requires the following steps \citep{epr1935,bohm1951,bell1964,brunner2014,chsh1969}. One firstly considers a composite physical entity ${S}_{12}$, prepared in an initial state $p$ and such that the individual entities ${S}_1$ and ${S}_2$ can be recognized as component parts of ${S}_{12}$. Then, one performs the coincidence measurements $XY$, $X=A,A'$, $Y=B,B'$, where each $XY$ consists in performing the measurement $X$ on $S_1$, with possible outcomes $X_i$, $i=1,2$, and the measurement $Y$ on $S_2$, with possible outcomes $Y_j$, $j=1,2$. The component entities $S_1$ and $S_2$ have interacted in the past, but are spatially separated when the $XY$s are performed. If the outcomes $X_i$ and $Y_j$ can only be equal to $\pm 1$, the expected value of the coincidence measurement $XY$ is just the correlation function
\begin{equation} \label{correlationfunction}
E(X,Y)=\sum_{i,j=1}^{2}X_iY_j \mu(X_iY_j)=\mu(X_1Y_1)-\mu(X_1Y_2)-\mu(X_2Y_1)+\mu(X_2Y_2)
\end{equation}
where $\mu(X_iY_j)$ is the joint probability of obtaining the outcome $X_i$ in a measurement of $X$ on $S_1$ and $Y_j$ in a measurement of $Y$ on $S_2$. Next, one calculates the quantity 
\begin{equation} \label{chshfactor}
\Delta_{CHSH}=E(A',B')+E(A',B)+E(A,B')-E(A,B)
\end{equation}
called the `CHSH factor' , and inserts it into the CHSH inequality
\begin{equation} \label{chsh}
-2 \le \Delta_{CHSH} \le 2
\end{equation}
The inequality in Equation (\ref{chsh}) follows from an assumption of `local separability', or `local realism' \citep{epr1935,bell1964}, which is reasonable in classical physics. Equivalently, Equation (\ref{chsh}) follows from the requirement that a classical Kolmogorovian model of probability exists for the measured correlations between $S_1$ and $S_2$ \citep{accardifedullo1982,pitowsky1989}. We finally notice that the CHSH factor in Equation (\ref{chshfactor}) is mathematically bound by the values $-4$ and $+4$.

The standard formulation of quantum mechanics associates the entities $S_1$ and $S_2$ with the complex Hilbert spaces ${\mathscr H}_1$ and ${\mathscr H}_2$, respectively, hence the composite entity $S_{12}$ is associated with the tensor product Hilbert space ${\mathscr H_1}\otimes {\mathscr H}_2$. The possible (pure) states of $S_1$ and $S_2$ are represented by unit vectors of ${\mathscr H}_1$ and ${\mathscr H}_2$, respectively, and the measurements that can be performed on $S_1$ and $S_2$ are represented by self-adjoint operators on ${\mathscr H_1}$ and ${\mathscr H}_2$, respectively. The states of $S_{12}$ that are represented by product vectors of ${\mathscr H_1}\otimes {\mathscr H}_2$ are called `product states', while the states that cannot be represented by product vectors are called `entangled states'. Analogously, the measurements on $S_{12}$ that are represented by product self-adjoint operators on ${\mathscr H_1}\otimes {\mathscr H}_2$ are called `product measurements'. But, ${\mathscr H_1}\otimes {\mathscr H}_2$ also contains non-product self-adjoint operators, which thus represent `entangled measurements'. In an entangled measurement, `at least' one eigenvector of the corresponding self-adjoint operator represents an entangled state \citep{brunner2014}. In the Bell-type test above, ${\mathscr H}_1$ and ${\mathscr H}_2$ are both isomorphic to the complex Hilbert space $\mathbb{C}^2$ of all ordered pairs of complex numbers, hence ${\mathscr H_1}\otimes {\mathscr H}_2$ is isomorphic to the complex Hilbert space $\mathbb{C}^2 \otimes \mathbb{C}^2$. 

The CHSH inequality in Equation (\ref{chsh}) is violated in quantum mechanics, which is interpreted as due to the presence of entanglement between the entities $S_1$ and $S_2$ that are recognized as component parts of $S_{12}$. This entanglement is typically attributed to the initial state $p$ of the composite entity $S_{12}$ being the singlet spin state, i.e. a maximally entangled state, and the coincidence measurements $XY$ being product measurements, $X=A,A'$, $B'B'$. 

In the situation of the singlet spin state and product coincidence measurements:

(i) the CHSH factor is equal to $2\sqrt{2} \approx 2.83$ and is called `Cirel'son's bound' \citep{cirelson1980,cirelson1993}, as it is typically considered the maximum value reachable in quantum mechanics in the presence of product measurements;

(ii) the conditions that, for every $i,j=1,2$, $X,X'=A,A'$, $X' \ne X$, $Y,Y'=B,B'$, $Y'\ne Y$,
\begin{eqnarray}
\sum_{j=1,2} \mu(X_iY_j)&=&\sum_{j=1,2}\mu (X_iY'_{j}) \label{marginal1} \\
\sum_{i=1,2} \mu (X_iY_j)&=&\sum_{i=1,2}\mu (X'_iY_{j}) \label{marginal2}
\end{eqnarray}
are trivially satisfied and are called the `marginal law', or `no-signaling' \citep{brunner2014,peresterno2004,horodecki2009}, or `marginal selectivity' \citep{barros2015,dzhafarov2016}, conditions. A violation of Equations (\ref{marginal1}) and (\ref{marginal2}) is typically considered as uninteresting in physics, as it would entail the possibility of `signaling'. We will return to points (i) and (ii) in Section \ref{theoretical}.

Bell-type tests have been extensively performed in micro-physical domains, mainly to identify the phenomenon of `non-locality', and they all confirm the predictions of quantum mechanics (see, e.g., \citet{genovese2005,vienna2013,urbana2013}).

Coming to conceptual-linguistic domains, we performed several Bell-type tests in the form of both cognitive tests on human participants \citep{aertssozzo2011,aertssozzo2014,aertsarguellesbeltrangerientesozzo2023b,bertinietal2023,aertsleporinisozzo2025}, document retrieval tests on corpora of documents of English language \citep{beltrangeriente2019,aertsbeltrangerientesozzo2021}, and image retrieval tests on web search engines \citep{arguelles2018,arguellessozzo2020,bertinietal2023}. We particularly tested the conceptual combination {\it The Animal Acts}, meant as a composition of the individual conceptual entities {\it Animal} and {\it Acts}, where the term ``acts'' refers to the possible sounds that can be emitted by the animal, and {\it The Animal eats the Food}, meant as a composition of the individual conceptual entities {\it Animal} and {\it Food}. The idea was to perform measurements on the composite entity, i.e. the conceptual combination, which were the conceptual-linguistic analogue of the coincidence measurements described above, and test the CHSH inequality to identify the eventual presence of entanglement. We found a systematic violation of the CHSH inequality, in some cases beyond Cirel'son's bound, together with a systematic violation of the marginal law conditions.

While the violation of the CHSH inequality was in substantial agreement with the predictions of quantum mechanics, indicating a non-classical situation where entanglement occurs, as in physical domains, the additional violation of the marginal law conditions and Cirel'son's bound was somewhat unexpected, as they are not believed to occur in physical domains (see points (i) and (ii) above). These findings led us to start a theoretical analysis of a problem that is connected with entanglement and is usually overlooked in quantum mechanics, namely, the `identification problem', that is, the problem of identifying individual entities that are the component parts of a composite entity by performing measurements on the latter \citep{aertssozzo2014}. Thus, we elaborated a general theoretical framework to model any Bell-type test, independently of the nature, physical or conceptual-linguistic, of the entities involved, within the formalism of quantum mechanics in Hilbert space \citep{aertssozzo2014,aertsetal2019}. In this theoretical framework, one applies the quantum mechanical prescription that the composite entity $S_{12}$ is associated with a complex Hilbert space whose dimension is determined by the number of distinct outcomes of the measurements performed on $S_{12}$. In the case of a Bell-type test, each coincidence measurement $XY$, $X=A,A'$, $Y=B,B'$, has four distinct outcomes, hence $S_{12}$ should be associated with the Hilbert space $\mathbb{C}^{4}$ of all ordered 4-tuples of complex numbers. Only in the attempt of identifying two individual entities $S_1$ and $S_2$ as parts of $S_{12}$, one considers possible isomorphisms with the tensor product Hilbert space $\mathbb{C}^{2} \otimes \mathbb{C}^{2}$, where each copy of $\mathbb{C}^{2}$ takes into account the fact that measurements with two distinct outcomes can be performed on $S_1$ and $S_2$ in a Bell-type test. And it is only at the stage in which individual entities are identified from measurements performed on the composite entity that entanglement may arise. We proved that, in general, no unique isomorphism exists between $\mathbb{C}^{4}$ and $\mathbb{C}^{2} \otimes \mathbb{C}^{2}$, which is the reason why, from a mathematical point of view, different ways exist to account for entanglement being present within the composite entity $S_{12}$ with respect to the individual entities $S_1$ and $S_2$ that are identified as parts of $S_{12}$. 

To formalize these more general situations that appear in conceptual-linguistic domains, we needed a more rigorous characterization of the non-classical aspects of entanglement than the intuitive, but somewhat misleading, picture of entanglement that is typically provided in physical domains. As we will explain in Section \ref{theoretical}, we found this characterization in the fact that  
entanglement formalizes the non-classical situation in which the probabilities of a coincidence measurement on $S_{12}$ cannot be written as products of probabilities of measurements on the individual entities $S_1$ and $S_2$ that are identified as parts of $S_{12}$. Hence, entanglement is a `relational property' between the coincidence measurements and the measurements on the component entities. Only when the marginal law conditions in Equations (\ref{marginal1}) and (\ref{marginal2}) are satisfied in all coincidence measurements, the entanglement of these different coincidence measurements can be captured in the state of the composite entity. Indeed, if the marginal law conditions are satisfied in all coincidence measurements, then one can prove that a unique isomorphism exists which connects $\mathbb{C}^{4}$ with $\mathbb{C}^{2} \otimes \mathbb{C}^{2}$, in which case $\mathbb{C}^{4}$ can be directly identified with $\mathbb{C}^{2} \otimes \mathbb{C}^{2}$. This means that the situation typically reported in quantum mechanics, namely, entanglement as a consequence of an initial entangled state and product coincidence measurements, is not the general one \citep{aertssozzo2014}. In the general situation in which the marginal law conditions are empirically violated, no unique isomorphism exists between $\mathbb{C}^{4}$ and $\mathbb{C}^{2} \otimes \mathbb{C}^{2}$, hence entanglement cannot by captured only by the initial state. As a matter of fact, empirical violations of Equations (\ref{marginal1}) and (\ref{marginal2}) have been identified in linguistic-conceptual domains, as anticipated above, but also in physical domains \citep{adenierkhrennikov2007,adenierkhrennikov2016,bednorz2017,deraedt2012,kupczynski2017}. However, little attention has been devoted so far to the violation of the marginal law conditions in physical tests of entanglement, because the latter has been attributed to  artifacts of the measurement process \citep{adenierkhrennikov2016}. 

On the other side, the simultaneous violations of the CHSH inequality, the marginal law conditions and Cirel'son's bound led us to investigate in depth the identification problem above, reaching conclusions on entanglement as a phenomenon which diverge in some aspects from the typical tenet of quantum mechanics summarized at the beginning of this section. Illustrating this new theoretical perspective on entanglement will be the aim of Section \ref{theoretical}.

\section{A new theoretical perspective on entanglement\label{theoretical}}
Our general theoretical perspective on entanglement as a phenomenon appearing in both physical and conceptual-linguistic domains was motivated by the awareness that the phenomenon is `holistically deeper' than the intuitive picture typically provided by quantum physicists (see \citet{aertssozzo2014,aertsetal2019,aertsleporinisozzo2025}). This led us to reconsider the following elements. 

(i) When investigating if/how the phenomenon occurs for conceptual entities, which, differently from physical entities, cannot be localized in space, one has to distil the non-classical aspects of entanglement that are independent of the presence in space of the entities under study. As such, one needs to dissociate entanglement from the physical notion of `non-locality'. 

(ii) Because of (i), the typical relationship between the marginal law conditions and no-signaling has also to be carefully re-analyzed because it mainly relies on the intuitive, but in our opinion misleading, picture of two physical entities that exist separately and are localized in space, hence can exchange signals. 

(iii) Though in Bell-type tests on conceptual entities, one still considers composite entities that are bipartite, i.e. consist of two component entities, generally speaking, one does not generally know how the two component entities relate to each other and to the composite entity. 

Points (i)--(iii) led us to look for a mathematically more rigorous way to introduce entanglement that was closer to its underlying nature, independently of its physical or conceptual-linguistic declination. 

Let us now consider a bipartite entity, prepared in a defined state and such that coincidence measurements can be performed on it and probabilities can be defined as large number limits of relative frequencies of these coincidence measurements. We agree that entanglement is present if the probabilities obtained from the coincidence measurements on the bipartite entity cannot be factorized as the product of the probabilities obtained from the individual, or separate, measurements that compose the coincidence measurements. 

This characterization of entanglement incorporates its non-classical aspects because, if classical physical entities are separated from each other in space, then the probabilities above do factorize. Also in quantum mechanics, when a bipartite entity is in a product state and the coincidence measurements are represented by product operators, then the probabilities above factorize. This means that in a situation modeled within the quantum formalism, if the probabilities of the coincidence measurements do not factorize, at least one of these aspects, initial state or coincidence measurements, fails. The most studied failing case is when the state of the bipartite entity is not a product state, thus an entangled state. Little attention was paid to the situation where the coincidence measurements are not represented by product operators, which also entails a lack of factorization of probabilities. 

Let us then come to the relationship between entanglement and the violation of the CHSH inequality within our general perspective. We refer to Pitowsky's theory of correlation polytopes and their connection with classical Kolmogorovian probabilities \citep{pitowsky1989}. In this theory, the CHSH inequality is equivalent to the existence of a classical Kolmogorovian probability model for the corresponding probabilities. Hence, a violation of the CHSH inequality is sufficient for the presence of a non-classical, possibly quantum, probability model. 

Little attention has been devoted so far to the situation where entanglement has its origin in the violation of both the CHSH inequality and the marginal law conditions. The reason is that the violation of the marginal law conditions, which are also called the no-signaling conditions (see Section \ref{empirical}), is typically interpreted as indicating the presence of signaling, hence it is marked as trivially uninteresting by physicists. We believe that this is not necessarily true, because the marginal law conditions only constitute a `sufficient', but `not necessary', condition for no-signaling. Hence, a violation of the marginal law conditions does not entail in itself the presence of signaling. On the other side, we have studied in detail examples of Bell-type situations where a violation of the CHSH inequality occurs together with a violation of the marginal law conditions \citep{aerts1982,aerts1983,aerts1991,aerts2005}, which allowed us to conclude that the violation of the marginal law conditions is due to a lack of some form of symmetry, rather than a consequence of the existence of signaling. In these examples, the violation of the CHSH inequality was interpreted as a genuine expression of entanglement provoked by the presence of `potential correlations which are only actualized when the coincidence measurements are performed', which already there provided an explanation for the appearance of entanglement which was more palatable than the typical `spooky action-at-a-distance view'.  

That the violation of the marginal law conditions is only a secondary effect, due to a lack of enough symmetry in the bipartite entity under study, and not a primary effect responsible of signaling, is also evident from the fact that in some of our previous tests the CHSH inequality is violated by an amount that exceeds Cirel'son's bound, whereas this is not the case for micro-physical entities that are entangled. We can understand the condition for the existence of this bound when we consider Cirel'son's proof of it in detail. We can then note that it is necessary to be able to represent the four considered coincidence measurements present in the CHSH inequality expression by one self-adjoint operator in the considered Hilbert space. This indicates the supposed presence of a very large and rigorous quantum coherence that incorporates the four coincidence measurements simultaneously and independently of how and when they may be performed separately. This very large internal coherence is not easy to bring present in conceptual-linguistic tests.

We have thus summarized the key points in which our general perspective on entanglement differs from the typical view of entanglement in physics. As mentioned above, there is one important point that arose from our research of entanglement in conceptual-linguistic domains, namely, a deeper holistic nature is present when two entities entangle each other. The entangled entity is more than usually supposed to be a new entity in itself only still connected to the original entities from which it was formed in a very specific way, and this has become even more clear to us by analysing more deeply how concepts are entangled in human language. As mentioned in Section \ref{intro}, we have recently introduced a phenomenon of `contextual updating' taking place in human language relative to the global meaning carried by the whole context, which contains an important part of our new understanding. That also the marginal law conditions are violated as a consequence of the entangled entity behaving fully as a new entity relative to the global meaning context is clearly identified in language, and it is also clearly seen that this is not related to the presence of signals. We believe that a similar process takes place when two physical entities become entangled, namely, a new entity is created that forms itself not primarily with respect to the two component entities, but contextually with respect to the global quantum coherence present, in which the two component entities will generally play an important role, but in principle not only they \citep{aertsarguellesbeltrangerientesozzo2023a} (see Section \ref{conclusion}). In this sense, we believe that the situation where both the CHSH inequality and the marginal law conditions, are violated is the default situation in terms of entanglement.

We finally come to the mathematical representation in Hilbert space of the general perspective on entanglement summarized in this section. In that regard, the analysis of the problem of identifying component entities from measurements performed on a composite bipartite entity (Section \ref{empirical}), led us to work out a quantum-theoretic framework to model any Bell-type test which violates both the marginal law conditions, the CHSH inequality and Cirel'son's bound. In this quantum-theoretic framework, we explicitly introduce entangled measurements \citep{aertssozzo2014,arguellessozzo2020,aertsbeltrangerientesozzo2021,aertsarguellesbeltrangerientesozzo2023b,aertsleporinisozzo2025}. Specifically, we have proved that, whenever two entities combine to form a composite bipartite entity, a strong form of entanglement is created between the individual entities composing it, which is such that, not only the state of the composite entity is entangled, but also the coincidence measurements are entangled.

We are now ready to present in detail the information retrieval tests we performed on Italian linguistic corpora with the aim of identifying conceptual entanglement. This will be the aim of Section \ref{tests}.

\section{Description of the tests on Italian linguistic corpora\label{tests}}
As anticipated in Section \ref{intro}, we performed three Bell-type information retrieval tests using selected corpora of Italian language, namely, the corpus ``CORIS/CODIS'', the corpus ``PAISÀ'', and the corpus ``Italian Web 2020''. Thus, let us preliminarily provide some information about these corpora.

The corpus ``CORIS/CODIS''\footnote{See the webpage \url{https://corpora.ficlit.unibo.it/coris_eng.html}} is a corpus of written Italian language, which has been publicly accessible online since September 2001. Initiated in 1998 by R. Rossini Favretti, the project aimed at developing a comprehensive and representative reference corpus of contemporary written Italian, designed for ease of access and use. The corpus, now containing 165 million words, is regularly updated every three years through an embedded monitoring corpus. It is composed of a collection of authentic and widely used electronic texts, carefully selected to reflect usage of current Italian language.

The corpus ``PAISÀ''\footnote{See the webpage \url{https://www.corpusitaliano.it/en/index.html}} is a large corpus of authentic contemporary Italian texts from the web. It was created within the project that holds the same name (PAISÀ is the acronym of ``Piattaforma per l’Apprendimento dell’Italiano Su corpora Annotati'', or platform for learning of Italian language on annotated corpora) with the aim to provide a large resource of freely available Italian texts for language learning by studying authentic text materials. The corpus has a dimension of around 250 million tokens and contains freely available and distributable web texts, collected in September/October 2010.

The corpus ``Italian Web 2020'', also known as ``ItTenTen20'',\footnote{See the webpage \url{https://www.sketchengine.eu/ittenten-italian-corpus/}} is part of the corpus manager ``Sketch Engine'' and, more specifically, it is the version for Italian language of the ``TenTen Corpus Family'', which includes more than 50 languages. All corpora in this family are prepared according to the same criteria. The Italian corpus, whose most recent version contains 12.4 billion words, is made up of texts collected from the web in November/December 2019 and December 2020. Its sample texts were checked manually and content with poor quality text was removed.

Let us now come to the description of the three Bell-type tests.

In the first test, we studied the Italian translation {\it L'Animale fa un Verso} of the conceptual combination {\it The Animal Acts}, considered as a composite entity made up of the conceptual entities {\it Animal} and {\it Acts}. In this test, we chose the Italian translations of the examples of animals and acts considered in our previous studies on {\it The Animal Acts} \citep{aertssozzo2011,aertssozzo2014,aertsarguellesbeltrangerientesozzo2023b,bertinietal2023,aertsleporinisozzo2025}, and used the corpora ``CORIS/CODIS'' and ``Italian Web 2020''. To set up a Bell-type test, we applied in each corpus the procedure illustrated in \citet{beltrangeriente2019,aertsbeltrangerientesozzo2021}, as follows.

For the coincidence measurement $AB$, we calculated the number of times that each of the following strings

$A_1B_1$: {\it Il Cavallo Ringhia} ({\it The Horse Growls})

$A_1B_2$: {\it Il Cavallo Nitrisce} ({\it The Horse Whinnies})

$A_2B_1$: {\it L'Orso Ringhia} ({\it The Bear Growls})

$A_2B_2$: {\it L'Orso Nitrisce} ({\it The Bear Whinnies})

appeared in the texts of the corpus.

For the coincidence measurement $AB'$, we calculated the number of times that each of the following strings

$A_1B'_1$: {\it Il Cavallo Sbuffa} ({\it The Horse Snorts})

$A_1B'_2$: {\it Il Cavallo Miagola} ({\it The Horse Meows})

$A_2B'_1$: {\it L'Orso Sbuffa} ({\it The Bear Snorts})

$A_2B'_2$: {\it L'Orso Miagola} ({\it The Bear Meows})

appeared in the texts of the corpus.

For the coincidence measurement $A'B$, we calculated the number of times that each of the following strings

$A'_1B_1$: {\it La Tigre Ringhia} ({\it The Tiger Growls})

$A'_1B_2$: {\it La Tigre Nitrisce} ({\it The Tiger Whinnies})

$A'_2B_1$: {\it Il Gatto Ringhia} ({\it The Cat Growls})

$A'_2B_2$: {\it Il Gatto Nitrisce} ({\it The Cat Whinnies})

appeared in the texts of the corpus.

For the coincidence experiment $A'B'$, we calculated the number of times that each of the following strings 

$A'_1B'_1$: {\it La Tigre Sbuffa}  ({\it The Tiger Snorts})
 
$A'_1B'_2$: {\it La Tigre Miagola}  ({\it The Tiger Meows})

$A'_2B'_1$: {\it Il Gatto Sbuffa} ({\it The Cat Snorts})

$A'_2B'_2$: {\it Il Gatto Miagola} ({\it The Cat Meows})

appeared in the texts of the corpus.

All retrieved strings, or pairs of lemmas, in the jargon of computational linguistics, were inspected manually to avoid false positive cases. Then, for each coincidence measurement $XY$, $X=A,A'$, $Y=B,B'$, we calculated the relative frequency of appearance of the string $X_iY_j$, $i,j=1,2$, which we considered, in the large number limit, as the probability of appearance $\mu(X_iY_j)$ or, equivalently as the probability that the outcome $X_iY_j=\pm 1$ is obtained in the coincidence measurement $XY$. Table \ref{tab1} reports the probabilities of appearance computed in this way. In it, we indicate the outcomes in English language, for the sake of clarity.

Next, the probabilities of appearance $\mu(X_iY_j)$, $i,j=1,2$, were used to calculate the expectation values, or correlation functions, $E(X,Y)$, using Equation (\ref{correlationfunction}), $X=A,A'$, $Y=B,B'$. Finally, we calculated the CHSH factor in Equation (\ref{chshfactor}) and compared it with the CHSH inequality in Equation (\ref{chsh}).

\begin{table}
\begin{center}
\begin{tabular}{|c | c | c | c|}
\hline
\multicolumn{4}{|c|}{Test 1: {\it The Animal Acts} original} \\
\hline
\hline
Example & Probability  & CORIS/CODIS & Italian Web 2020 \\
\hline
\multicolumn{4}{|c|}{Measurement $AB$} \\
\hline
{\it The Horse Growls} & $\mu(A_1B_1)$ & 0 & 0.013 \\
{\it The Horse Whinnies} & $\mu(A_1B_2)$ & 1 & 0.9508 \\
{\it The Bear Growls} & $\mu(A_2B_1)$ & 0 & 0.0364 \\ 
{\it The Bear Whinnies} & $\mu(A_2B_2)$ & 0 & 0 \\ 
\hline
\multicolumn{2}{|c|}{Expectation value $E(AB)$} & --1 & --0.9744 \\
\hline
\multicolumn{4}{|c|}{Measurement $AB'$} \\
\hline
{\it The Horse Snorts} & $\mu(A_1B'_1)$ & 1 & 0.9358 \\
{\it The Horse Meows} & $\mu(A_1B'_2)$ & 0 & 0 \\
{\it The Bear Snorts} & $\mu(A_2B'_1)$ & 0 & 0.064 \\ 
{\it The Bear Meows} & $\mu(A_2B'_2)$ & 0 & 0 \\ 
\hline
\multicolumn{2}{|c|}{Expectation value $E(AB')$} & 1 & 0.8716 \\
\hline 
\multicolumn{4}{|c|}{Measurement $A'B$} \\
\hline
{\it The Tiger Growls} & $\mu(A'_1B_1)$ & 0.5 & 0.1948 \\
{\it The Tiger Whinnies} & $\mu(A'_1B_2)$ & 0 & 0 \\
{\it The Cat Growls} & $\mu(A'_2B_1)$ & 0.5 & 0.8052 \\ 
{\it The Cat Whinnies} & $\mu(A'_2B_2)$ & 0 & 0 \\ 
\hline
\multicolumn{2}{|c|}{Expectation value $E(A'B)$} & 0 & --0.6104 \\
\hline
\multicolumn{4}{|c|}{Measurement $A'B'$} \\
\hline
{\it The Tiger Snorts} & $\mu(A'_1B'_1)$ & 0 & 0.0006 \\
{\it The Tiger Meows} & $\mu(A'_1B'_2)$ & 0 & 0.0032 \\
{\it The Cat Snorts} & $\mu(A'_2B'_1)$ & 0 & 0.0173 \\ 
{\it The Cat Meows} & $\mu(A'_2B'_2)$ & 1 & 0.9788 \\ 
\hline
\multicolumn{2}{|c|}{Expectation value $E(A'B')$} & 1 & 0.9590 \\
\hline 
\hline
\multicolumn{2}{|c|}{CHSH factor $\Delta_{CHSH}$} & 3 & 2.1946 \\ 
\hline
\end{tabular}
\end{center}	
\caption{We report in comparison for the considered corpora the statistical data of the first test on the Italian translation of the conceptual combination {\it The Animal Acts} where we used the same examples of animals and acts as in the previous tests on this combination \citep{aertssozzo2011,aertsarguellesbeltrangerientesozzo2023b,bertinietal2023,aertsleporinisozzo2025}. In both corpora ``CORIS/CODIS'' and ``Italial Web 2020'', the CHSH factor exceeds the value of $2$, indicating a violation of the CHSH inequality, which substantially agrees with previous results on English corpora. \label{tab1}}
\end{table}
As we can see from Table \ref{tab1}, in both corpora ``CORIS/CODIS'' and ``Italian Web 2020'', the CHSH factor exceeds the numerical value of $2$, which indicates a violation of the CHSH inequality in Equation (\ref{chsh}), hence a `deviation from classicality' and, because of our considerations in Sections \ref{empirical} and \ref{theoretical}, the presence of `quantum entanglement' between the component entities {\it Animal} and {\it Acts}. In the first case, relative to the corpus ``CORIS/CODIS'', the CHSH factor is equal to $3$, thus Cirel'son's bound is violated too. This result shows substantial agreement with the information retrieval tests in \citet{beltrangeriente2019,aertsbeltrangerientesozzo2021,bertinietal2023}, where corpora of English language were employed, and also with the cognitive tests in \citet{aertsarguellesbeltrangerientesozzo2023b,bertinietal2023,aertsleporinisozzo2025}, where human participants were employed. In the second case, relative to the corpus ``Italian Web 2020'', the CHSH factor is equal to $2.19$, which shows substantial agreement with the cognitive tests in \citet{aertssozzo2011,aertssozzo2014}.

It should be noted that the corpus ``PAISÀ'' does not have the capability to systematically detect entanglement, due to an insufficient number of occurrences of the utilized strings. In addition, we did not retrieve in our searches too many entries with the other two corpora either. This is why we decided to perform a second test on the Italian translation {\it L'Animale fa un Verso} of the conceptual combination {\it The Animal Acts}, which was a variant of the original test. We decided to identify additional animals and associated sound emissions, i.e. acts, that could provide a sufficient number of occurrences to perform the test, replicating the methodology of the original test. Concerning the latter, indeed, we noticed that, though horses are well-known in Italian culture, and their characteristic neigh and snort are commonly recognized, there were insufficient occurrences of horses whinning and snorting in our corpora. Consequently, we replaced the horse with a more common domestic animal, the dog, which is familiar across many cultures. We selected barking as the primary vocalization and howling as the secondary one for the dog. The bear, which can growl, was substituted by the rooster, which sings, due to its greater familiarity and prevalence. The tiger, present in some Italian zoos and numerous films, was replaced by the wolf as a wild animal due to the wolf's comparable recognition. Then, we repeated the test using the corpora ``CORIS/CODIS'', ``PAISÀ'', and ``Italian Web 2020''. More specifically, we considered a maximum of 9 words between the two lemmas and then, again, removed false positives from the extraction. The results are presented in Table \ref{tab2}, where we indicate the outcomes in both Italian and English language, for the sake of completeness.

\begin{table}
\small
\begin{center}
\begin{tabular}{|c | c | c | c | c|}
\hline
\multicolumn{5}{|c|}{Test 2: {\it The Animal Acts} modified} \\
\hline
\hline
Example & Probability & CORIS/CODIS & PAISÀ & Italian Web 2020 \\
\hline
\multicolumn{5}{|c|}{Measurement $AB$} \\
\hline
{\it Il Cane Canta} ({\it The Dog Sings}) & $\mu(A_1B_1)$ & 0.007 & 0.024 & 0.0370 \\
{\it Il Cane Abbaia} ({\it The Dog Barks}) & $\mu(A_1B_2)$ & 0.9932 & 0.9762 & 0.9364 \\
{\it Il Lupo Canta} ({\it The Wolf Sings}) & $\mu(A_2B_1)$ & 0 & 0 & 0.0133 \\
{\it Il Lupo Abbaia} ({\it The Wolf Barks}) & $\mu(A_2B_2)$ & 0 & 0 & 0.0132 \\
\hline
\multicolumn{2}{|c|}{Expectation value $E(AB)$} & --0.9865 & --0.9524 &  --0.8996 \\
\hline
\multicolumn{5}{|c|}{Measurement $AB'$} \\
\hline
{\it Il Cane Ulula} ({\it The Dog Howls}) & $\mu(A_1B'_1)$ & 0.6250 & 0.4286 & 0.4550 \\
{\it Il Cane Miagola} ({\it The Dog Meows}) & $\mu(A_1B'_2)$ & 0 & 0 & 0.0226 \\
{\it Il Lupo Ulula} ({\it The Wolf Howls}) & $\mu(A_2B'_1)$ & 0.3750 & 0.5714 & 0.5217 \\
{\it Il Lupo Miagola} ({\it The Wolf Meows}) & $\mu(A_2B'_2)$ & 0 & 0 & 0.0006 \\
\hline
\multicolumn{2}{|c|}{Expectation value $E(AB')$} & 0.2500 & --0.1429 & --0.0887 \\
\hline 
\multicolumn{5}{|c|}{Measurement $A'B$} \\
\hline
{\it Il Gallo Canta} ({\it The Rooster Sings}) & $\mu(A'_1B_1)$ & 0.9818 & 1 & 0.9534 \\
{\it Il Gallo Abbaia} ({\it The Rooster Barks}) & $\mu(A'_1B_2)$ & 0 & 0 & 0.0012 \\
{\it Il Gatto Canta} ({\it The Cat Sings}) & $\mu(A'_2B_1)$ & 0.0182 & 0 & 0.0268 \\
{\it Il Gatto Abbaia} ({\it The Cat Barks}) & $\mu(A'_2B_2)$  & 0 & 0 & 0.0186 \\
\hline
\multicolumn{2}{|c|}{Expectation value $E(A'B)$} & 0.9636 & 1 & 0.9441 \\
\hline
\multicolumn{5}{|c|}{Measurement $A'B'$} \\
\hline
{\it Il Gallo Ulula} ({\it The Rooster Howls}) & $\mu(A'_1B'_1)$ & 0 & 0 & 0.0006 \\
{\it Il Gallo Miagola} ({\it The Rooster Meows}) & $\mu(A'_1B'_2)$ & 0 & 0 & 0 \\
{\it Il Gatto Ulula} ({\it The Cat Howls}) & $\mu(A'_2B'_1)$ & 0 & 0 & 0.0078 \\
{\it Il Gatto Miagola} ({\it The Cat Meows}) & $\mu(A'_2B'_2)$ & 1 & 1 & 0.9916 \\
\hline
\multicolumn{2}{|c|}{Expectation value $E(A'B')$} & 1 & 1 & 0.9844 \\
\hline 
\hline
\multicolumn{2}{|c|}{CHSH factor $\Delta_{CHSH}$} & 3.2001 & 2.8095 & 2.7393 \\ 
\hline
\end{tabular}
\end{center}
		
\caption{We report in comparison for the considered corpora the statistical data of the second test on the Italian translation of the conceptual combination {\it The Animal Acts} where we used new examples of animals and acts. Also in this case, in both corpora ``CORIS/CODIS'', ``PAISÀ'', and ``Italial Web 2020'', the CHSH factor exceeds the value of $2$, again indicating a violation of the CHSH inequality, in substantial agreement with previous results on English corpora. \label{tab2}}
\end{table}

As we can see from Table \ref{tab2}, in all corpora ``CORIS/CODIS'', ``PAISÀ'', and ``Italian Web 2020'', the CHSH factor exceeds the numerical value of $2$, which indicates a violation of the CHSH inequality in Equation (\ref{chsh}). In addition, the CHSH factor oscillates around Cirel'son's bound. In all cases, however, a strong deviation from classicality occurs, which indicates that entanglement is again at play in the conceptual combination {\it The Animal Acts}. 

Finally, in the third test, we investigated the Italian translation {\it L'Animale mangia il Cibo} of the conceptual combination {\it The Animal eats the Food}, which we considered as a composite entity made up of the individual conceptual entities {\it Animal} and {\it Food}. We used the same examples of animals and food as in \citet{aertsarguellesbeltrangerientesozzo2023b} and worked on the corpus ``Italian Web 2020'', where we considered a maximum distance of 4 words between the subject and the verb, and a maximum distance of 5 words between the verb and the object. Furthermore, we included the most common synonyms of the word ``eat'', such as ``nourish'' and ``feed''. To set up a Bell-type test, we applied to the procedure illustrated above, as follows.

For the coincidence measurement $AB$, we calculated the number of occurrences of the following strings

$A_1B_1$: {\it Il Gatto mangia l'Erba} ({\it The Cat eats the Grass})

$A_1B_2$: {\it Il Gatto mangia la Carne} ({\it The Cat eats the Meat})

$A_2B_1$: {\it La Mucca mangia l'Erba} ({\it The Cow eats the Grass})

$A_2B_2$: {\it La Mucca mangia la Carne} ({\it The Cow eats the Meat}).

For the coincidence measurement $AB'$, we calculated the number of occurrences of the following strings

$A_1B'_1$: {\it Il Gatto mangia il Pesce} ({\it The Cat eats the Fish})

$A_1B'_2$: {\it Il Gatto mangia le Noci} ({\it The Cat eats the Nuts})

$A_2B'_1$:  {\it La Mucca mangia il Pesce} ({\it The Cow eats the Fish})

$A_2B'_2$:  {\it La Mucca mangia le Noci} ({\it The Cow eats the Nuts}).

For the coincidence measurement $A'B$, we calculated the number of occurrences of the following strings

$A'_1B_1$: {\it Il Cavallo mangia l'Erba} ({\it The Horse eats the Grass})

$A'_1B_2$: {\it Il Cavallo mangia la Carne} ({\it The Horse eats the Meat})

$A'_2B_1$: {\it Lo Scoiattolo mangia l'Erba} ({\it The Squirrel eats the Grass})

$A'_2B_2$: {\it Lo Scoiattolo mangia la Carne} ({\it The Squirrel eats the Meat}).

For the coincidence experiment $A'B'$, we calculated the number of occurrences of the following strings

$A'_1B'_1$: {\it Il Cavallo mangia il Pesce} ({\it The Horse eats the Fish})
 
$A'_1B'_2$: {\it Il Cavallo mangia le Noci} ({\it The Horse eats the Nuts})

$A'_2B'_1$: {\it Lo Scoiattolo mangia il Pesce} ({\it The Squirrel eats the Fish})

$A'_2B'_2$: {\it Lo Scoiattolo mangia le Noci} ({\it The Squirrel eats the Nuts}).

Also in this case, all entries were inspected manually to avoid false positives. Then, for each coincidence measurement $XY$, $X=A,A'$, $Y=B,B'$, we calculated the relative frequency of appearance of the string $X_iY_j$, $i,j=1,2$, which we considered, in the large number limit, as the probability of occurrence $\mu(X_iY_j)$, i.e. the probability that the outcome $X_iY_j=\pm 1$ is obtained in $XY$. Table \ref{tab3} reports the probabilities of appearance computed in this way. We again indicate the outcomes in English language, for the sake of clarity.

\begin{table}
\begin{center}
\begin{tabular}{|c|c|c|}
\hline
\multicolumn{3}{|c|}{Test 3: {\it The Animal eats the Food}} \\
\hline
\hline
Example & Probability & Italian Web 2020 \\
\hline
\multicolumn{3}{|c|}{Measurement $AB$} \\
\hline
{\it The Cat eats the Grass} & $\mu(A_1B_1)$ & 0.1710 \\
{\it The Cat eats the Meat}  & $\mu(A_1B_2)$ & 0.2684 \\
{\it The Cow eats the Grass}  & $\mu(A_2B_1)$ & 0.5467 \\ 
{\it The Cow eats the Meat}  & $\mu(A_2B_2)$ & 0.0139 \\ 
\hline
\multicolumn{2}{|c|}{Expectation value $E(AB)$} & --0.6302 \\
\hline
\multicolumn{3}{|c|}{Measurement $AB'$} \\
\hline
{\it The Cat eats the Fish} & $\mu(A_1B'_1)$ & 1 \\
{\it The Cat eats the Nuts} & $\mu(A_1B'_2)$ & 0 \\
{\it The Cow eats the Fish} & $\mu(A_2B'_1)$ & 0 \\ 
{\it The Cow eats the Nuts} & $\mu(A_2B'_2)$ & 0 \\ 
\hline
\multicolumn{2}{|c|}{Expectation value $E(AB')$} & 1 \\
\hline 
\multicolumn{3}{|c|}{Measurement $A'B$} \\
\hline
{\it The Horse eats the Grass} & $\mu(A'_1B_1)$ & 0.7526 \\
{\it The Horse eats the Meat} & $\mu(A'_1B_2)$ & 0.2371 \\
{\it The Squirrel eats the Grass} & $\mu(A'_2B_1)$ &  0.0103 \\ 
{\it The Squirrel eats the Meat} & $\mu(A'_2B_2)$ &  0 \\ 
\hline
\multicolumn{2}{|c|}{Expectation value $E(A'B)$} & 0.5052 \\
\hline
\multicolumn{3}{|c|}{Measurement $A'B'$} \\
\hline
{\it The Horse eats the Fish} & $\mu(A'_1B'_1)$ & 0.5 \\
{\it The Horse eats the Nuts} & $\mu(A'_1B'_2)$ & 0 \\
{\it The Squirrel eats the Fish} & $\mu(A'_2B'_1)$ & 0 \\ 
{\it The Squirrel eats the Nuts} & $\mu(A'_2B'_2)$ & 0.5 \\ 
\hline
\multicolumn{2}{|c|}{Expectation value $E(A'B')$} & 1 \\
\hline 
\hline
\multicolumn{2}{|c|}{CHSH factor $\Delta_{CHSH}$} & 3.1354 \\ 
\hline
\end{tabular}
\end{center}
		
\caption{We report in comparison for the considered corpora the statistical data of the first test on the Italian translation of the conceptual combination {\it The Animal eats the Food} where we used the same examples of animals and acts as in the previous tests on this combination \citep{aertssozzo2011,aertsarguellesbeltrangerientesozzo2023b,bertinietal2023,aertsleporinisozzo2025}. In both corpora ``CORIS/CODIS'' and ``Italial Web 2020'', the CHSH factor exceeds the value of $2$, indicating a violation of the CHSH inequality, which substantially agrees with previous results on English corpora. \label{tab3}}
\end{table}

As we can see from Table \ref{tab3}, the CHSH factor in Equation (\ref{chshfactor}) violates both the CHSH inequality in Equation (\ref{chsh}) and Cirel'son's bound, in substantial agreement with the empirical patterns identified in \citet{aertsarguellesbeltrangerientesozzo2023b}. Also in this case, the violation of the CHSH inequality indicates the presence of entanglement in the conceptual combination {\it The Animal eats the Food}.

Summing up, the three information retrieval Bell-type tests we have analyzed in this section lead one to draw the preliminary conclusion that a violation of the CHSH inequality systematically occurs. This suggests that the phenomenon of entanglement is systematically present whenever two (or more) concepts combine, and the phenomenon is independent of the language, English or Italian, that is used to reveal it. This means that we have identified here a deep conceptual structure underlying human language and, more generally, human cognition \citep{aertsetal2019}. This conclusion will be sustained by the elaboration of an explicit quantum-theoretic model for the data presented in this section, as we will in Section \ref{model}.

\section{A quantum model in Hilbert space\label{model}}
We elaborate a quantum mathematical model in Hilbert space for the data presented in Section \ref{tests}, limiting ourselves to the corpus ``Italian Web 2020'', since the latter enabled to collect data for all tests. The model here rests on the general quantum-theoretic framework that we have worked out for the modeling of any Bell-type situation, independently of the nature, physical or conceptual-linguistic, of the entities involved \citep{aertssozzo2014,aertsetal2019,aertsbeltrangerientesozzo2021,aertsarguellesbeltrangerientesozzo2023b}.

As already done in Section \ref{tests}, though the tests were conducted in Italian language, we will use the English language translations of the conceptual entities and the examples of animals, acts and food considered in the tests, for the sake of clarity.

In our quantum-theoretic framework, the conceptual entity under study ({\it The Animal Acts} or {\it The Animal eats the Food}) is a composite bipartite entity $S_{12}$ made up of the individual entities $S_1$ and $S_2$ ({\it Animal} and {\it Acts} or {\it Animal} and {\it Food}, respectively). We assume that $S_{12}$ is in an initial state $p$, which corresponds to the situation of an animal that makes a sound (in {\it The Animal Acts} case) or an animal that eats some food (in {\it The Animal eats the Food} case). Then, for every $X=A,A'$, $Y=B,B'$, the coincidence measurement $XY$ has four possible outcomes  $X_iY_j$, $i,j=1,2$, which correspond to the four strings, or pairs of lemmas, we looked at in $XY$ (see Tables \ref{tab1} and \ref{tab2} for {\it The Animal Acts} and Table \ref{tab3} for {\it The Animals eats the Food}). Next, each outcome $X_iY_j$ of $XY$ is associated with an outcome state, or eigenstate, $p_{X_iY_j}$, which corresponds to an example of an animal that makes a specific sound (in {\it The Animal Acts}, e.g., {\it The Bear Growls}, {\it The Dog Barks}) or an example of an animal that eats some specific food (in {\it The Animal eats the Food}, e.g., {\it The Cat eats the Fish}, {\it The Squirrel eats the Nuts}). Finally, the probability of appearance $\mu(X_iY_j)$ corresponds to the probability $P_{p}(X_iY_j)$ that the outcome $X_iY_j$ is obtained when the coincidence measurement $XY$ is performed on the composite entity $S_{12}$ in the initial state $p$.

Let us now come to the Hilbert space representation of the empirical notions identified above. In this regard, since each coincidence measurement $XY$, $X=A,A'$, $Y=B,B'$, has four outcomes, the composite entity $S_{12}$, meant as a `single entity', is associated with the complex Hilbert space $\mathbb{C}^{4}$ of all ordered 4-tuples of complex numbers. Moreover, each state $p$ of the composite entity is represented by a unit vector of $\mathbb{C}^{4}$ and each coincidence measurement on $S_{12}$ is represented by a self-adjoint operator or, equivalently, by a spectral family, on $\mathbb{C}^{4}$. On the other side, each outcome $X_iY_j$, $i,j=1,2$, is obtained by juxtaposing the outcomes $X_i$ and $Y_j$ (e.g., in {\it The Animal Acts}, the string ``horse whinnies'' is obtained by juxtaposing the lemmas ``horse'' and ``whinnies''). This defines a 2-outcome measurement $X$, $X=A,A'$, on the individual entity $S_1$ and a 2-outcome measurement $Y$, $Y=B,B'$, on the individual entity $S_2$. Since each of these individual entities is associated with the complex Hilbert space $\mathbb{C}^{2}$ of all ordered pairs of complex numbers, $S_{12}$, meant as a `composition of $S_1$ and $S_2$', is associated with the tensor product Hilbert space $\mathbb{C}^{2} \otimes \mathbb{C}^{2}$ (see Section \ref{empirical}).

The vector spaces $\mathbb{C}^{4}$ and $\mathbb{C}^{2} \otimes \mathbb{C}^{2}$ are formally isomorphic, where each isomorphism maps an orthonormal (ON) basis of $\mathbb{C}^{4}$ onto an ON basis of $\mathbb{C}^{2} \otimes \mathbb{C}^{2}$. The states of $S_{12}$ are represented by unit vectors of $\mathbb{C}^{4}$, which correspond, through the isomorphism, to vectors of $\mathbb{C}^{2} \otimes \mathbb{C}^{2}$, hence to either vectors that represent product states or vectors that represent entangled states. Analogously, the vector space $L(\mathbb{C}^{4})$ of all linear operators on $\mathbb{C}^{4}$ is isomorphic to the tensor product vector space $L(\mathbb{C}^{2}) \otimes L(\mathbb{C}^{2})$, where $L(\mathbb{C}^{2})$ is the vector space of all linear operators on $\mathbb{C}^{2}$. The measurements on $S_{12}$ are represented by self-adjoint operators, which correspond, through the isomorphism, to self-adjoint operators of $L(\mathbb{C}^{2}) \otimes L(\mathbb{C}^{2})$, hence to either self-adjoint operators that represent product measurements or self-adjoint operators that represent entangled measurements (see again Section \ref{empirical}). More precisely, let $I: \mathbb{C}^{4} \longrightarrow \mathbb{C}^{2} \otimes \mathbb{C}^{2}$ be an isomorphism mapping a given ON basis of $\mathbb{C}^{4}$ onto a given ON basis of $\mathbb{C}^{2} \otimes \mathbb{C}^{2}$. We say that a state $p$, represented by the unit vector $|p\rangle \in {\mathbb C}^4$, is a `product state with respect to $I$', if two states $p_A$ and $p_B$, represented by the unit vectors $|p_A\rangle \in {\mathbb C}^2$ and $|p_B\rangle \in {\mathbb C}^2$, respectively, exist such that $I|p\rangle=|p_A\rangle\otimes|p_B\rangle$. Otherwise, $p$ is an `entangled state with respect to $I$'. Analogously, we say that a measurement $e$, represented by the self-adjoint operator ${\mathscr E}$ on ${\mathbb C}^4$, is a `product measurement with respect to $I$', if two measurements $e_X$ and $e_Y$, represented by the self-adjoint operators  ${\mathscr E}_X$ and ${\mathscr E}_Y$, respectively, on ${\mathbb C}^2$ exist such that $I{\mathscr E}I^{-1}={\mathscr E}_X \otimes {\mathscr E}_Y$. Otherwise, $e$ is an `entangled measurement with respect to $I$'. Thus, the notion of entanglement does depend on the `isomorphism that is used to identify individual entities within a given composite entity'. 

With reference to a Bell-type setting, one can then prove the following statements \citep{aertssozzo2014}:

(i) if the coincidence measurements ${XY}$ and  ${XY'}$, $X=A,A'$, $Y,Y'=B,B'$, $Y \ne Y'$, are product measurements with respect to the isomorphism $I$, then, for every state $p$ of the composed entity $S_{12}$, the marginal law condition in Equation (\ref{marginal1}) is satisfied; 

(ii) if the coincidence measurements $XY$ and $X'Y$, $X,X'=A,A'$, $Y=B,B'$, $X \ne X'$, are product measurements with respect to the isomorphism $I$, then, for every state $p$ of $S_{12}$, the marginal law condition in Equation (\ref{marginal2}) is satisfied; 

(iii) if the marginal law conditions are satisfied in all coincidence measurements, then a unique isomorphism exists, which can be chosen to be the identity operator.

It follows from (i)--(iii) that, if the marginal law conditions in Equations (\ref{marginal1}) and (\ref{marginal2}) are empirically violated, as it frequently occurs in Bell-type tests on conceptual-linguistic entities, and the three tests in Section \ref{tests} do not make an exception,\footnote{One can verify the violation of the marginal law conditions in Equations (\ref{marginal1}) and (\ref{marginal2}) by directly inspecting the statistical data on Tables \ref{tab1}--\ref{tab3}.} one cannot find a unique isomorphism $I: \mathbb{C}^{4}\longrightarrow\mathbb{C}^{2} \otimes \mathbb{C}^{2}$ such that all measurements are product measurements with respect to $I$. In these cases, one cannot attribute the violation of the CHSH inequality by concentrating all the entanglement of the state-measurement situation in the initial state and assuming that all coincidence measurements are product measurements, as one does in Bell-type tests on physical entities.\footnote{As anticipated in Section \ref{empirical}, there are reasons to believe that the marginal law conditions are also violated in Bell-type tests on physical entities, which indicates that entangled measurements are present in the physical domain too. However, the violation of the marginal law conditions in physical tests is not large, hence it has hardly been investigated in depth \citep{aertsetal2019}.} Thus, if we set an isomorphism $I: \mathbb{C}^{4}\longrightarrow\mathbb{C}^{2} \otimes \mathbb{C}^{2}$, it is likely that both the initial state and all coincidence measurements are entangled \citep{aertssozzo2014}.

Keeping in mind the considerations above, we associate the composite conceptual entity $S_{12}$ ({\it The Animal Acts} or {\it The Animal eats the Food})  with the complex Hilbert space $\mathbb{C}^4$. Then, let $(1,0,0,0)$, $(0,1,0,0)$, $(0,0,1,0)$ and $(0,0,0,1)\}$ be the unit vectors of the canonical ON basis of $\mathbb{C}^4$, and let $I:\mathbb{C}^4  \longrightarrow \mathbb{C}^2 \otimes \mathbb{C}^2$ be the isomorphism such that the canonical ON basis of $\mathbb{C}^4$ coincides with the ON basis of the tensor product Hilbert space $\mathbb{C}^2 \otimes \mathbb{C}^2$ made up of the unit vectors $(1,0)\otimes (1,0)$, $(1,0)\otimes (0,1)$, $(0,1)\otimes (1,0)$ and $(0,1)\otimes (0,1)$. 

In the ON bases above, a given state $q$ of the composite entity is represented by the unit vector $|q\rangle=(ae^{i \alpha}, be^{i \beta}, ce^{i \gamma}, de^{i \delta})$, $a,b,c,d \ge 0$, $a^2+b^2+c^2+d^2=1$, $\alpha$, $\beta$, $\gamma$, $\delta \in \Re$, where $\Re$ is the real line. One easily proves that $|q\rangle$ represents a product state if and only if 
\begin{equation} \label{entanglementcondition}
ade^{i(\alpha+\delta)}-bce^{i(\beta+\gamma)}=0
\end{equation}
Otherwise, $|q\rangle$ represents an entangled state. 

Next, in analogy with Bell-type tests on physical entities, we represent the initial state $p$ of the composite entity $S_{12}$ by the unit vector 
\begin{equation}
|p\rangle=\frac{1}{\sqrt{2}}(0,1,-1,0) \label{singlet}
\end{equation}
This is a reasonable choice which we have also systematically made in previous work on this subject. Indeed, the vector in Equation (\ref{singlet}) represents in quantum mechanics the maximally entangled state that corresponds to the singlet spin state and is rotationally invariant, i.e. does not privilege any ON basis representation. Shifting to the conceptual-linguistic case, this entangled state corresponds to the abstract situation of an animal that makes a sound, or an animal that eats some food, without privileging any specific example of the conceptual entities under study. In other words, the choice to represent the initial state by the unit vector in Equation (\ref{singlet}) corresponds to the situation in which the composite entity is open to any type of measurement involving any example of animals and sounds, or any example of animals and food.\footnote{Another possible choice would have been to represent the initial state of the composite entity by the unit vector $|\psi\rangle=\frac{|\phi\rangle}{|||\phi\rangle||}$, where $|\phi\rangle=\sum_{X,Y}|\phi_{XY}\rangle$ and $|\phi_{XY}\rangle=\sum_{i,j}\sqrt{\mu(X_iY_j)}|p_{X_iY_j}\rangle$ (see the discussion in \citet{aertsleporinisozzo2025}, Section 5). We will not deal with this technical aspect here, for the sake of brevity.} In addition, we try, in this way, to concentrate as much entanglement as possible in the initial state (see also the analysis in \citet{aertsetal2019}). 

Further, for every $X=A,A'$, $Y=B,B'$, we represent the coincidence measurement $XY$ by the spectral family constructed on the ON basis of the four eigenvectors $|p_{X_iY_j}\rangle$, which represent the eigenstates $p_{X_iY_j}$ introduced above, where we set, for every $i,j=1,2$,
\begin{equation}
|p_{X_iY_j}\rangle=(a_{X_iY_j}e^{i \alpha_{X_iY_j}}, b_{X_iY_j}e^{i \beta_{X_iY_j}}, c_{X_iY_j}e^{i \gamma_{X_iY_j}}, d_{X_iY_j}e^{i \delta_{X_iY_j}}) \label{eigenvectors}
\end{equation}
The coefficients are such that $a_{X_iY_j}, b_{X_iY_j}, c_{X_iY_j}, d_{X_iY_j} \ge 0$ and $\alpha_{X_iY_j}, \beta_{X_iY_j},\gamma_{X_iY_j}, \delta_{X_iY_j} \in\Re$. We set $\alpha_{X_iY_j}=\beta_{X_iY_j}=\gamma_{X_iY_j}=\delta_{X_iY_j}=\theta_{X_iY_j}$, where $\theta_{X_iY_j}\in \Re$, for the sake of simplicity. One easily verifies that, for every $X=A,A'$, $Y=B,B'$, $XY$ is a product measurement if and only if all $|p_{X_iY_j}\rangle$s are product vectors. Otherwise, $XY$ is an entangled measurement.

Finally, for every $X=A,A'$, $Y=B,B'$, $i,j=1,2$, the probability $P_{p}(X_iY_j)$ of obtaining the outcome $X_iY_j$ in a measurement of ${XY}$ on the composite entity $S_{12}$ in the state $p$ is given by Born's rule of quantum mechanics, that is, 
\begin{equation} \label{bornrule}
P_{p}(X_iY_j)=|\langle p_{X_iY_j}|p\rangle|^2
\end{equation}

To find a quantum mathematical modeling of the data in Tables \ref{tab1}--\ref{tab3}, for every measurement $e_{XY}$, the four unit vectors in Equation  (\ref{eigenvectors}) have to satisfy the following sets of conditions.

(1) Normalization. The eigenvectors in Equation (\ref{eigenvectors})  are unit vectors, that is, for every $X=A,A'$, $Y=B,B'$, $i,j=1,2$,
\begin{equation}
a_{X_iY_j}^2+b_{X_iY_j}^2+c_{X_iY_j}^2+d_{X_iY_j}^2=1
\end{equation}

(2) Orthogonality. The eigenvectors in Equation (\ref{eigenvectors})  are mutually orthogonal, that is,  for every $X=A,A'$, $Y=B,B'$, $i,i',j,j'=1,2$, $i \ne i'$, $j \ne j'$, 
\begin{eqnarray}
\langle p_{X_{i}Y_{j}}|p_{X_{i'}Y_{j'}} \rangle&=&0 \\
\langle p_{X_{i}Y_{j}}|p_{X_{i}Y_{j'}} \rangle&=&0 \\
\langle p_{X_{i}Y_{j}}|p_{X_{i}Y_{j'}} \rangle&=&0
\end{eqnarray}

 (3) Empirical adequacy. For every $X=A,A'$, $Y=B,B'$, $i,j=1,2$, the probability $P_{p}(X_iY_j)$ coincides with the empirical probability $\mu(X_iY_j)$ in Tables \ref{tab1}--\ref{tab3}, that is,
\begin{equation}
|\langle p_{X_iY_j}|p\rangle|^{2}=\mu(X_iY_j)
\end{equation}
where we have used Born's rule in Equation (\ref{bornrule}).

We now provide an explicit solution, starting by the first test on {\it The Animal Acts} original. 

The eigenstates of the measurement ${AB}$ are represented by the unit vectors
\begin{eqnarray}
|p_{A_1B_1}\rangle &=&e^{i 182.68^{\circ}}(0.88,0.24,0.40,0) \label{Gsol_HG} \\
|p_{A_1B_2}\rangle &=&e^{i 0.39^{\circ}}(0,0.85,-0.53,0) \label{Gsol_HW} \\
|p_{A_2B_1}\rangle &=&e^{i 308.03^{\circ}}(0.47,-0.47,-0.74,0) \label{Gsol_BG} \\
|p_{A_2B_2}\rangle &=&e^{i 188.97^{\circ}}(0,0,0,-1) \label{Gsol_BW}
\end{eqnarray}
By applying the entanglement condition in Equation (\ref{entanglementcondition}), we find that ${AB}$ is an entangled measurement. In addition, Equation (\ref{entanglementcondition}) shows a larger deviation from zero in the unit vector $|p_{A_1B_2}\rangle$, which thus represents a relatively more entangled state. Indeed, the eigenstate $p_{A_1B_2}$ corresponds to {\it The Horse Whinnies}, which is more characteristic of the global meaning carried by {\it The Animal Acts}. Analogously, the unit vector $|p_{A_2B_2}\rangle$ represents a product state. Indeed, the eigenstate $p_{A_2B_2}$ corresponds to {\it The Bear Whinnies}, which is less characteristic of the global meaning carried by {\it The Animal Acts}.

The eigenstates of the measurement ${AB'}$ are represented by the unit vectors
\begin{eqnarray}
|p_{A_1B'_1}\rangle &=&e^{i 0.07^{\circ}}(0,0.50,-0.86,0)  \label{Gsol_HS} \\
|p_{A_1B'_2}\rangle &=&e^{i 10.64^{\circ}}(0,0,0,-1) \label{Gsol_HM} \\
|p_{A_2B'_1}\rangle &=&e^{i 81.84^{\circ}}(0,0.86,0.50,0) \label{Gsol_BS} \\
|p_{A_2B'_2}\rangle &=&e^{i 162.03^{\circ}}(1,0,0,0) \label{Gsol_BM}
\end{eqnarray}
Also in this case, ${AB'}$ is an entangled measurement. In addition, Equation (\ref{entanglementcondition}) shows a larger deviation from zero in the unit vector $|p_{A_1B'_1}\rangle$, which thus represents a relatively more entangled state. Indeed, the eigenstate $p_{A_1B'_1}$ corresponds to {\it The Horse Snorts}, which is more characteristic of the global meaning carried by {\it The Animal Acts}. Analogously, the unit vectors $|p_{A_1B'_2}\rangle$ and $|p_{A_2B'_2}\rangle$ represent product states. Indeed, the eigenstates $p_{A_1B'_2}$ and $p_{A_2B'_2}$ correspond to {\it The Horse Meows} and {\it The Bear Meows}, which are less characteristic of the global meaning carried by {\it The Animal Acts}.

The eigenstates of the measurement ${A'B}$ are represented by the unit vectors
\begin{eqnarray}
|p_{A'_1B_1}\rangle &=&e^{i 3.38^{\circ}}(-0.05,-0.32,-0.94,0)\label{Gsol_TG} \\
|p_{A'_1B_2}\rangle &=&e^{i 174.93^{\circ}}(0.99,-0.04,-0.04,0) \label{Gsol_TW} \\
|p_{A'_2B_1}\rangle &=&e^{i 245.49^{\circ}}(0.03,0.95,-0.32,0) \label{Gsol_CG} \\
|p_{A'_2B_2}\rangle &=&e^{i 239.45^{\circ}}(0,0,0,-1) \label{Gsol_CW}
\end{eqnarray}
Again, ${A'B}$ is an entangled measurement. In addition, Equation (\ref{entanglementcondition}) shows a larger deviation from zero in the unit vector $|p_{A'_2B_1}\rangle$, which thus represents a relatively more entangled state. Indeed, the eigenstate $p_{A'_2B_1}$ corresponds to {\it The Cat Growls}, which is more characteristic of the global meaning carried by {\it The Animal Acts}. Analogously, the unit vector $|p_{A'_2B_2}\rangle$ represents a product state. Indeed, the eigenstate $p_{A'_2B_2}$ corresponds to {\it The Cat Whinnies}, which are less characteristic of the global meaning carried by {\it The Animal Acts}.

Finally, the eigenstates of the measurement ${A'B'}$ are represented by the unit vectors
\begin{eqnarray}
|p_{A'_1B'_1}\rangle &=&e^{i 92.68^{\circ}}(0.47,-0.64,-0.60,0) \label{Gsol_TS} \\
|p_{A'_1B'_2}\rangle &=&e^{i 20.16^{\circ}}(0.14,0.09,0.01,-0.98)\label{Gsol_TM} \\
|p_{A'_2B'_1}\rangle &=&e^{i 0.69^{\circ}}(0.86,0.23,0.42,0.16) \label{Gsol_CS} \\
|p_{A'_2B'_2}\rangle &=&e^{i 138.27^{\circ}}(0.12,0.72,-0.67,0.08) \label{Gsol_CM}
\end{eqnarray}
Also ${A'B'}$ is an entangled measurement. In addition, Equation (\ref{entanglementcondition}) shows a larger deviation from zero in the unit vector $|p_{A'_2B'_2}\rangle$, which thus represents a relatively more entangled state. Indeed, the eigenstate $p_{A'_2B'_2}$ corresponds to {\it The Cat Meows}, which is more characteristic of the global meaning carried by {\it The Animal Acts}. Analogously, Equation (\ref{entanglementcondition}) shows a smaller deviation from zero in the unit vector $|p_{A'_2B'_1}\rangle$, which thus represents a relatively less entangled state. Indeed, the eigenstate $p_{A'_2B'_1}$ corresponds to {\it The Cat Snorts}, which is less characteristic of the global meaning carried by {\it The Animal Acts}.

The second test on {\it The Animal Acts} modified allows one to draw exactly the same considerations as the first one. Indeed, the eigenstates corresponding to examples whose meaning is closer to the global meaning context provided by {\it The Animal Acts} are represented by relatively more entangled states, namely, the eigenstates corresponding to  {\it The Dog Barks} in the measurement $AB$, {\it The Wolf Howls} in the measurement $AB'$, {\it The Rooster Sings} in the measurement $A'B$, and {\it The Cat Meows} in the measurement $A'B'$. Analogously, the eigenstates corresponding to examples which are farther than the global meaning context provided by {\it The Animal Acts} are represented by relatively less entangled states, namely, the eigenstates corresponding to  {\it The Wolf Barks} in the measurement $AB$, {\it The Wolf Meows} in the measurement $AB'$, {\it The Rooster Barks} in the measurement $A'B$, and {\it The Rooster Meows} in the measurement $A'B'$. For the sake of completeness, we report the complete set of eigenvectors in the following.
\begin{eqnarray}
|p_{A_1B_1}\rangle &=&e^{i 263.83^{\circ}}(0.61,-0.68,-0.41,0) \\
|p_{A_1B_2}\rangle &=&e^{i 32.84^{\circ}}(0,0.52,-0.85,0.08)  \\
|p_{A_2B_1}\rangle &=&e^{i 223.44^{\circ}}(0.19,0,-0.97,0.77) \\
|p_{A_2B_2}\rangle &=&e^{i 359.86^{\circ}}(0.77,0.49,0.33,0.23) \\
|p_{A_1B'_1}\rangle &=&e^{i 0.05^{\circ}}(-0.24,0.13,-0.83,-0.49) \\
|p_{A_1B'_2}\rangle &=&e^{i 277.68^{\circ}}(-0.88,0.34,0.13,0.31)  \\
|p_{A_2B'_1}\rangle &=&e^{i 214.46^{\circ}}(0.41,0.81,-0.22,0.37) \\
|p_{A_2B'_2}\rangle &=&e^{i 69.78^{\circ}}(0,0.47,0.50,-0.73) \\
|p_{A'_1B_1}\rangle &=&e^{i 58.96^{\circ}}(-0.03,0.55,-0.83,0.08) \\
|p_{A'_1B_2}\rangle &=&e^{i 246.26^{\circ}}(0.96,-0.23,-0.18,0)  \\
|p_{A'_2B_1}\rangle &=&e^{i 99.94^{\circ}}(0.28,0.75,0.51,0.32) \\
|p_{A'_2B_2}\rangle &=&e^{i 240.58^{\circ}}(0.09,0.30,0.10,-0.95) \\
|p_{A'_1B'_1}\rangle &=&e^{i 252.43^{\circ}}(0.12,0.18,0.15,-0.96) \\
|p_{A'_1B'_2}\rangle &=&e^{i 101.50^{\circ}}(0.94,-0.25,-0.25,0.03)  \\
|p_{A'_2B'_1}\rangle &=&e^{i 90.47^{\circ}}(0.33,0.70,0.58,0.26) \\
|p_{A'_2B'_2}\rangle &=&e^{i 9.14^{\circ}}(-0.03,0.64,-0.77,0) 
\end{eqnarray}

The third test on {\it The Animal eats the Food} completely aligns with the first two tests. Indeed, the eigenstates corresponding to examples whose meaning is closer to the global meaning of {\it The Animal eats the Food}, namely, the eigenstates corresponding to  {\it The Cow eats the Grass} in the measurement $AB$, {\it The Cat eats the Fish} in the measurement $AB'$, {\it The Horse eats the Grass} in the measurement $A'B$, and {\it The Squirrel eats the Nuts} in the measurement $A'B'$. Analogously, the eigenstates corresponding to examples that are farther than {\it The Animal eats the Food} are represented by relatively less entangled states, namely, the eigenstates corresponding to  {\it The Cow eats the Meat} in the measurement $AB$, {\it The Cow eat the Nuts} in the measurement $AB'$, {\it The Squirrel eats the Meat} in the measurement $A'B$, and {\it The Horse eats the Meat} in the measurement $A'B'$. Also in this case, we report the complete set of eigenvectors, as follows.
\begin{eqnarray}
|p_{A_1B_1}\rangle &=&e^{i 160.28^{\circ}}(0.48,0.25,0.84,0) \\
|p_{A_1B_2}\rangle &=&e^{i 0.37^{\circ}}(-0.57,0.80,0.09,0.67)  \\
|p_{A_2B_1}\rangle &=&e^{i 247.95^{\circ}}(0.67,0.51,-0.54,0.09) \\
|p_{A_2B_2}\rangle &=&e^{i 157.84^{\circ}}(0.01,0.12,-0.04,-0.99) \\
|p_{A_1B'_1}\rangle &=&e^{i 44.11^{\circ}}(0,0.71,-0.71,0) \\
|p_{A_1B'_2}\rangle &=&e^{i 2.18^{\circ}}(0.37,0.66,0.66,0)  \\
|p_{A_2B'_1}\rangle &=&e^{i 261.61^{\circ}}(0.93,-0.26,-0.26,0) \\
|p_{A_2B'_2}\rangle &=&e^{i 120.36^{\circ}}(0,0,0,-1) \\
|p_{A'_1B_1}\rangle &=&e^{i 60.74^{\circ}}(0,0.26,-0.97,0) \\
|p_{A'_1B_2}\rangle &=&e^{i 2.26^{\circ}}(0.20,0.95,0.26,0)  \\
|p_{A'_2B_1}\rangle &=&e^{i 334.89^{\circ}}(0.98,-0.19,-0.05,0) \\
|p_{A'_2B_2}\rangle &=&e^{i 150.72^{\circ}}(0,0,0,-1) \\
|p_{A'_1B'_1}\rangle &=&e^{i 0.05^{\circ}}(-0.15,0.12,-0.88,-0.43) \\
|p_{A'_1B'_2}\rangle &=&e^{i 275.00^{\circ}}(-0.97,0.12,0.12,0.14)  \\
|p_{A'_2B'_1}\rangle &=&e^{i 245.71^{\circ}}(0,0.44,0.44,-0.78) \\
|p_{A'_2B'_2}\rangle &=&e^{i 66.72^{\circ}}(0.15,0.88,-0.12,0.43) 
\end{eqnarray}

The quantum mathematical model for the data presented in Section \ref{tests} is thus completed and allows us to make some interesting considerations about the appearance of entanglement in conceptual-linguistic domains, as follows. 

(a) The results obtained on corpora of Italian language confirm and strengthen those obtained in both cognitive tests on human participants \citep{aertssozzo2014,aertsarguellesbeltrangerientesozzo2023b,aertsleporinisozzo2025} and information retrieval tests on corpora of English language \citep{arguellessozzo2020,aertsbeltrangerientesozzo2021}. In particular, the individual concepts {\it Animal} and {\it Acts} entangle when they combine to form the combination {\it The Animal Acts}. Similarly, the individual concepts {\it Animal} and {\it Food} entangle when combined to form the combination {\it The Animal eats the Food}. This is because the concepts {\it Animal} ({\it Animal}) and {it Acts} ({\it Food}) carry `meaning', but {\it The Animal Acts} ({\it The Animal eats the Food}) also carries its own meaning, which is not attributed to the latter by separately attributing meaning to {\it Animal} ({\it Animal}) an {\it Acts} ({\it Food}). On the contrary, the combination process `creates additional meaning' in a way that violates the rules of classical compositional semantics. We believe that this process of `meaning attribution and creation as a result of conceptual combination can be exactly captured by the quantum phenomenon of entanglement. Or, equivalently, meaning is the conceptual-linguistic counterpart of what entanglement is in physical domains.

(b) In both {\it The Animal Acts} and {\it The Animal eats the Food} cases, the empirical violation of the marginal law conditions of Kolmogorovian probability forbids concentrating all the entanglement of the state-measurement situation in the initial state of the composite entity, as we have seen above. As discussed in Section \ref{theoretical}, this violation cannot be interpreted as a clue that signaling is at place. On the contrary, we believe that the violation of the marginal law conditions indicates that we are in the presence of a stronger form of entanglement than the one typically identified in physical domains, and this stronger form of entanglement is the one that is most frequent, even in Bell-type tests on physical entities. In this sense, the violation of the CHSH inequality beyond Cirel'son's bound in some of the tests we performed should not come as a surprise, but it is again a consequence of this stronger form of entanglement involving both states and measurements.

(c) In the quantum mathematical modeling above, most of the eigenstates of the coincidence measurements are entangled too. This can be explained by observing that, in each coincidence measurement, all possible outcomes correspond themselves to combinations of concepts, e.g., the outcome {\it The Wolf Howls} is itself a combination of the concepts {\it Wolf} and {\it Howls}. Analogously, the outcome {\it The Cat eats the Fish} is itself a combination of the concepts {\it Cat} and {\it Fish}. Thus, on the basis of the theoretical connections above between entanglement and meaning, we believe that a non-classical process of meaning attribution and creation occurs at the level of examples too. In addition, in each coincidence measurement, some eigenstates exhibit a relatively higher degree of entanglement than others. Again, this can be explained by the fact that entanglement captures meaning, hence eigenstates with higher degree of entanglement correspond to examples whose meaning is closer to the overall meaning carried by the relative composite entity.

\section{Conclusion\label{conclusion}}
We conclude this article by inquiring more deeply into point (a) at the end of the previous section on the process of meaning attribution and creation which occurs in the combination process of two concepts. We indeed believe that this process is the conceptual-linguistic counterpart of the process of entanglement formation in physical domains. This would shed new light on the nature of quantum entanglement, which is thus independent of the domain, conceptual-linguistic or physical, where entanglement occurs.

As we have seen at the end of Section \ref{model}, the conceptual combination {\it The Animal Acts} ({\it The Animal eats the Food}) is attributed meaning as a whole entity and not by separately attributing meaning to the individual concepts {\it Animal} and {\it Acts} ({\it Animal} and {\it Food}).  
This means that `the combination of two concepts is generally an emergent process in which the new meaning that is created depends on the whole context that is relevant to the combination'. This is more evident if, instead of considering the combination of two concepts, we consider a text produced by human language and study the relationship between its overall meaning and the meaning of all the concepts that correspond to the words appearing in the text. Each time a new word, hence a new concept, is added to the text, the overall meaning of the latter `contextually updates', and this updating continues to occur until all words are added and the text is completed. From the point of view of the mathematical representation, this process of meaning attribution and creation, which we have called `contextual updating' in previous work \citep{aertsarguellesbeltrangerientesozzo2023b,aertsarguellesbeltrangerientesozzo2023a}, has to correspond to the formation of an entangled state, because this is exactly how entangled states are formed in the tensor product Hilbert space. Equivalently, entangled states accomplish the contextual updating in the mathematical formalism of quantum mechanics. Coming back to the combination of two concepts, the concept {\it Animal} ({\it Animal}) is an abstraction of all possible animals and the concept {\it Acts} ({\it Food}) is an abstraction of all possible sounds produced by animals (food eaten by animals). But, the meaning of {\it The Animal Acts} ({\it The Animal eats the Food}) is not created by separately considering abstractions of animals (animals) and abstractions of acts (food) and then combining these abstractions. On the contrary, abstractions of animals making a sound (animals eating some food) are directly considered, and this occurs in a coherent way that is represented by a superposed, specifically, entangled, state. This is exactly the way we refer to when we speak about `contextual updating'.

To conclude, we believe that a similar and more specific process of contextual updating occurs in each coincidence measurement at the level of eigenstate selection. Indeed, we have seen in Section \ref{model}, point (c), that, in each coincidence measurement, the eigenstate with highest probability of appearance is relatively more entangled than the others. We have explained this by observing that these entangled states are more characteristic, that is, closer in meaning, to the meaning context provided by the corresponding composite entity. Let us consider, e.g., the coincidence measurement $AB$ for {\it The Animal Acts}. That the example with highest probability of appearance, {\it The Horse Whinnies}, is also more entangled than the other examples can be attributed to a subtle process of contextual updating, namely, {\it The Horse Whinnies} is closer to the global meaning context provided by {\it The Animal Acts}, together with {\it Horse}, {\it Bear}, {\it Growls} and {\it Whinnies}.\footnote{This conclusion agrees with some findings on the entanglement between the measured entity and the measuring apparatus in the measurement process that we have recently obtained in an investigation on the appearance of quantum structures in the so-called `warping effect of human perception' \citep{aertsarguellessozzo2025}.}

As anticipated in Section \ref{empirical}, we have recently put forward the hypothesis that the process of contextual updating is at the basis of entanglement formation in physics too, which suggests that `entanglement as contextual updating' provides a better explanation than the `spooky action at a distance' phenomenon of how entanglement is produced in physics \citep{aertsarguellesbeltrangerientesozzo2023a}.

\section*{Acknowledgments}
This work was supported by the project ``New Methodologies for Information Access and Retrieval with Applications to the Digital Humanities'', scientist in charge S. Sozzo, financed within the fund ``DIUM -- Department of Excellence 2023--27'' and by the funds that remained at the Vrije Universiteit Brussel at the completion of the ``QUARTZ (Quantum Information Access and Retrieval Theory)'' project, part of the ``European Union Marie Sklodowska-Curie Innovative Training Network 721321'', with Diederik Aerts as principle investigator for the Brussels part of the network.

\end{document}